\documentclass{ifacconf}

\usepackage{graphicx}      
\usepackage{natbib}        
\usepackage{amsmath, calc}
\usepackage{amsfonts}
\usepackage{xcolor}
\newtheorem{Rem}{Remark}
\newtheorem{Ass}{Assumption}
\newtheorem{Corollary}{Corollary}
\usepackage{algorithm}
\usepackage{algpseudocode}
\usepackage{comment}
\begin{document}
\begin{frontmatter}

\title{Sample-Efficient Model-Free Policy Gradient Methods for Stochastic LQR via Robust Linear Regression\thanksref{footnoteinfo}} 

\thanks[footnoteinfo]{Bowen Song acknowledges the support of the International Max Planck Research School for Intelligent Systems (IMPRS-IS).}

\author[First]{Bowen Song} 
\author[Second]{Sebastien Gros} 
\author[First]{Andrea Iannelli}

\address[First]{Institute for Systems Theory and Automatic Control, University of Stuttgart, Germany (e-mail: bowen.song,andrea.iannelli@ist.uni-stuttgart.de)}
\address[Second]{Department of Engineering Cybernetics,
Norwegian University of Science and Technology 7491 Trondheim, Norway  (e-mail: sebastien.gros@ntnu.no)}

\begin{abstract}                
Policy gradient algorithms are widely used in reinforcement learning and belong to the class of approximate dynamic programming methods. This paper studies two key policy gradient algorithms—the Natural Policy Gradient and the Gauss–Newton Method—for solving the linear quadratic regulator problem for unknown systems using stochastic data. The main challenge is the bias of the random gradient estimator when one employs least-squares due to the errors-in-variables setting. This issue is addressed by proposing a robust primal–dual estimation procedure. Using this improved gradient estimation scheme, this paper delivers an estimator with a convergence rate of order $\mathcal{O}(\epsilon^{-1})$. Theoretical results are further supported by numerical experiments.
\end{abstract}


\end{frontmatter}

\section{Introduction}
Policy gradient methods (PGMs) are fundamental tools in reinforcement learning [\cite{bertsekas2019reinforcement}], as they aim to optimize a parameterized policy with respect to the performance objective. Understanding their convergence properties is important [\cite{cen2023globalconvergencepolicygradient}], as it significantly impacts the reliability of deploying PGMs in real-world applications [\cite{ma2024stochasticonlineoptimizationcyberphysical}].

The linear quadratic regulator (LQR) problem serves as a canonical benchmark for studying the convergence behavior of PGMs in continuous state and action spaces [\cite{doi:10.1137/20M1382386,doi:10.1137/23M1554771}]. In the seminal work by \cite{pmlr-v80-fazel18a}, three representative PGMs were analyzed for solving the LQR problem: policy gradient descent (PGD), natural policy gradient (NPG), and the Gauss–Newton method (GNM). They all require knowledge of the cost gradient, which explicitly depends on the system dynamics. However, in many practical settings, the system matrices are unknown or inaccessible, making direct gradient computation infeasible. A natural extension is therefore to address the case of unknown dynamics by estimating the system matrices from data. The works by \cite{zhao2025policygradientadaptivecontrol, SongIannelli+2025+398+412} investigate this framework under bounded-noise settings.



In parallel, there are direct approaches that perform policy optimization without explicitly identifying the system dynamics. Following \cite{pmlr-v80-fazel18a}, which used a zeroth-order method to estimate gradients from noise-free data for implementing PGD/NPG method, several subsequent studies extended the PGD framework to more general settings, for example, stochastic noise by \cite{doi:10.1137/20M1382386}. In these analyses, the sample complexity required to obtain an $\epsilon$-optimal policy was initially shown to be $\mathcal{O}(\epsilon^{-4})$. 
By using stochastic approximation, \cite{pmlr-v89-malik19a} reduced the sample complexity of PGD to $\mathcal{O}(\epsilon^{-2})$, and it was further improved to $\mathcal{O}(\epsilon^{-1})$ in \cite{JMLR:v26:24-0636}.
Beyond the two categories discussed above, there exists another class of methods that neither estimate the entire system matrix via online identification nor rely on zeroth-order gradient estimation. Instead, these approaches directly estimate the specific matrix blocks required to construct a policy gradient update. This idea was first introduced in \cite{pmlr-v80-tu18a}. The work \cite{9511623} applied this approach to deterministic linear systems using GNM. Later, \cite{9691800} extended it to stochastic GNM without convergence guarantees under noisy data. More recently, \cite{JMLR:v24:22-0644} and \cite{doi:10.1137/23M1554771} applied the NPG method to stochastic systems and established convergence guarantees with a sample complexity $\mathcal{O}(\epsilon^{-1}(\ln\epsilon^{-1})^2)$ using ergodic data, and $\mathcal{O}(\epsilon^{-1}(\ln\epsilon^{-1})^7)$ using online data, respectively.

In this work, we apply the natural policy gradient (NPG) and Gauss–Newton method (GNM) to solve the linear quadratic regulator (LQR) problem for an unknown linear system subject to stochastic noise. To construct the quantities required by NPG and GNM, we employ a primal–dual estimation scheme that provides consistent estimates from noisy data. This scheme can be viewed as a general framework for analyzing linear regression problems with errors-in-variables. To further accelerate estimation, we introduce a multi-epoch refinement procedure, which improves the statistical rate and achieves a sample complexity of $\mathcal{O}(\epsilon^{-1})$. Once the matrix estimates are obtained, we perform NPG and GNM updates and establish global convergence guarantees for both algorithms. Compared with \cite{JMLR:v24:22-0644,doi:10.1137/23M1554771}, which apply only NPG method and require collecting new data at every policy update, our method reuses a single dataset across all iterations, which is key to achieving better sample complexity. To the best of the authors’ knowledge, this is the lowest sample complexity achieved so far for applying NPG and GNM to the LQR problem.

\subsection*{Notation}
We denote $A\succeq 0$ and $A\succ0$ as positive semidefinite and positive definite symmetric matrices, respectively. The Kronecker product is represented as $\otimes$ , $vec(A)=[a_1^\top,a_2^\top,...,a_n^\top]^\top$ stacks the columns of matrix $A$ into a vector, $vecv(v)=[v_1^2,v_1v_2,...,v_1v_n,v_2^2,...,v_2v_n,...,v_n^2]^\top$ rearranges the entries of vector $v$ in this specific pattern, $vecs(P)=[p_{11},2p_{12},...,2p_{1n},p_{22},...,2p_{2n},...,p_{nn}]^\top$ stacks the upper-triangular part of matrix $P\succ0$. For matrices, $\lVert \cdot\rVert_F$, $\lVert \cdot\rVert$ denote respectively their Frobenius norm, induced $2$-norm. $I_n$ is the identity matrix with $n$ row/columns. The symbol $\lambda_i(A)$ denotes the smallest $i$-th eigenvalue of the square matrix $A$. The symbol $\lceil x \rceil$ denotes the ceil function which returns the smallest integer greater or equal than $x\in\mathbb{R}$.
\section{Problem Setting and Preliminaries}



In this work, we consider the following averaged infinite-horizon LQR problem, where the system is subject to additive stochastic noise:
\begin{equation}\label{setting}
\begin{split}
        \min_{u_t}& \lim_{T\rightarrow +\infty}\frac{1}{T}\mathop{\mathbb{E}}_{x_0, w_t} \sum_{t=0}^{T-1}\left( x_t^\top Q x_t +u_t ^\top R u_t\right),\\
        &s.t. ~ x_{t+1}=Ax_t+Bu_t+w_t, \\
        &x_0 \sim \mathcal{N}(0,\Sigma_0),~w_t\sim \mathcal{N}(0,\Sigma_w),
\end{split}
\end{equation}
where $A\in \mathbb{R}^{n_x \times n_x}$, $B\in \mathbb{R}^{n_x \times n_u}$ denote the unknown but stabilizable system matrices. The covariance matrices satisfy $\Sigma_0,\Sigma_w\succ 0$ and $Q,R\succ 0$ are the weight matrices. We define the set of stabilizing feedback gains as:
\begin{equation*}
    \mathcal{S} := \left\{ K \in \mathbb{R}^{n_u \times n_x} \,\big|\, A_K := A + B K \text{ is Schur stable} \right\}.
\end{equation*}
The infinite-horizon average cost under a linear state-feedback policy $u_t = K x_t$ with $K \in \mathcal{S}$ is defined as:
\begin{equation}\label{CostFunction}
    C(K) := \lim_{T \to \infty} \frac{1}{T} \mathbb{E}_{x_0, w_t} \bigg[ \sum_{t=0}^{T-1} x_t^\top Q_K x_t \bigg],
\end{equation}
with $Q_K := Q + K^\top R K$. For any stabilizing policy $K \in \mathcal{S}$, the gradient of the cost function $C(K)$ is given by:
\begin{equation}\label{Gradient}
    \nabla C(K) = 2 E_K \Sigma_K,
\end{equation}
where $E_K := \left( R + B^\top P_K B \right) K + B^\top P_K A$; $P_K$ is the unique solution to the equation $P_K=A_K^\top P_K A_K+Q_K$, and $\Sigma_K$ is the average covariance matrix associated with $K\in \mathcal{S}$, defined as
\begin{equation}\label{definedAverage}
    \Sigma_K:=\lim_{T\rightarrow +\infty}\frac{1}{T}\sum_{t=0}^{T-1} \Sigma_t,~\text{with} ~\Sigma_t:={\mathbb{E}}_{x_0,w_t}[x_t x_t^\top].
\end{equation}
It is a well-known fact (\cite{lewis2012optimal}) that the optimal $K^*$ minimizing $C(K)$ satisfies
\begin{subequations}
    \begin{align}
K^*&=-(R+B^\top P_{K^*}B)^{-1}B^\top P_{K^*}A, \label{Kpolicyimprovement} \\
P_{K^*}&=Q+A^\top P_{K^*}A \\ \notag
&\quad-A^\top P_{K^*}B(R+B^\top P_{K^*}B)^{-1}B^\top P_{K^*}A.
    \end{align}
\end{subequations}
The average covariance matrix associated with the optimal $K^*$ is denoted as $\Sigma_{K^*}$. 

Following \cite{pmlr-v80-fazel18a}, it is now well established that the function $C$ is \emph{gradient-dominated} and \emph{almost-smooth}. These properties form the basis for establishing convergence guarantees of policy gradient methods. For brevity, we do not restate the corresponding definitions here; detailed formulations can be found in \cite[Lemma 1 and Lemma 2]{Full}. For later use, however, we recall the local Lipschitz continuity property of the cost function $C$.
\begin{lem} \cite[Lemma 4]{Full} \label{Lipschitz}   Suppose $K',K \in \mathcal{S}$ are such that:
    \begin{equation}\label{SigmaK1}
    \lVert K-K' \rVert \leq \min\bigg\{h(C(K)) ,\lVert K\rVert\bigg\},
\end{equation}
with $h(C(K)):=\frac{\lambda_1(\Sigma_w)\lambda_1(Q)}{4C(K)\lVert B \rVert(\lVert A\rVert+\lVert B \rVert b_K(C(K))+1)}$, it holds that:
   \begin{equation}
       \lVert C(K)-C(K')\rVert \leq h_{C}(C(K))\lVert K-K' \rVert,
   \end{equation}
where 
      \begin{equation*}\label{Pertubation1}
    \begin{split}
    h_{C}(C(K)) := 6 \big( \frac{C(K)}{\lambda_1(\Sigma_w)\lambda_1(Q)}\big)^2\big( b^2_{K}(C(K))\lVert R \rVert\lVert B \rVert(\lVert A\rVert\\+\lVert B \rVert b_{K}(C(K)))+b_{K}(C(K))\lVert R\rVert\big)\mathrm{Tr}(\Sigma_w).
    \end{split}
\end{equation*} and $b_K(C(K))$ is an upper bound on $\lVert K \rVert$ given by: 
\begin{equation*}\label{boundK}
    \begin{split}
        &b_{K}(C(K)):=\frac{1}{\lambda_1(R) } \big( \lVert B \rVert \lVert A\rVert\frac{C(K)}{\lambda_1(\Sigma_w)}\\
        &+\sqrt{{(C(K)-C(K^*))\big(\lVert R\rVert+\lVert B \rVert^2 \frac{C(K)}{\lambda_1(\Sigma_w)}\big)}{\lambda_1(\Sigma_w)^{-1}}}\big).  
    \end{split}
\end{equation*}
\end{lem}
\subsection{Convergence Analysis of Policy Gradient Methods}
In this subsection, we revisit the convergence properties of model-based policy gradient methods, namely the Natural Policy Gradient (NPG) and the Gauss–Newton Method (GNM). Their update rules for all $i \in \mathbb{Z}_+$ are given by:
\begin{itemize}
\item {Natural Policy Gradient:} 
\begin{equation}\label{updateNPG}
\begin{split}
    &K_{i+1} = K_i - \eta \nabla C(K_i) \Sigma_{K_i}^{-1}\\
    &=K_i-2\eta \left[( R + B^\top P_{K_i} B ) K_i + B^\top P_{K_i} A\right];
\end{split}
\end{equation}
\item {Gauss-Newton Method:} 
\begin{equation}
    \begin{split}\label{updateGMN}
        &K_{i+1} = K_i - \eta (R + B^\top P_{K_i} B)^{-1} \nabla C(K_i)\Sigma_{K_i}^{-1}\\
          &=K_i-2\eta\left[K_i+(R+B^\top P_{K_i}B)^{-1}B^\top P_{K_i}A\right], 
    \end{split}
\end{equation}
\end{itemize}
where $\eta > 0$ denotes the step size.
The detailed derivations of these update rules is given in \cite{Full}. 
\begin{thm}\cite[Theorem 2]{Full}\label{NPGTheorem}
     Suppose the initial $K_0 \in \mathcal{S}$, and consider the natural policy gradient iteration \eqref{updateNPG}, with $\eta \leq \frac{1}{2\lVert R +B^\top P_{K_0}B\rVert}$.
Then
\begin{equation}\label{PGConvergence}
    C(K_{i+1})-C(K^*) \leq \gamma_N (C(K_{i})-C(K^*)),~\forall i\in \mathbb{Z}_+,
\end{equation}
with $\gamma_N:= 1-\frac{2\eta \lambda_1{(R)} \lambda_1(\Sigma_w)}{\lVert \Sigma_{K^*}\rVert}$. For $\eta = \frac{1}{2\lVert R +B^\top P_{K_0}B\rVert}$ and given any accuracy gap $\epsilon >0$, if the number of iterations $n \geq \frac{\lVert \Sigma_{K^*}\rVert}{2\eta\lambda_1(\Sigma_w)\lambda_1{(R)}}\log \frac{C(K_{0})-C(K^*)}{\epsilon},$
then $C(K_n)-C(K^*) \leq \epsilon.$
\end{thm}
\begin{thm}\cite[Theorem 3]{Full}\label{GNTheorem}
        Suppose the initial $K_0 \in \mathcal{S}$, and consider the Gauss-Newton iteration \eqref{updateGMN}, with $\eta \leq \frac{1}{2}$. Then
    \begin{equation}
    C(K_{i+1})-C(K^*) \leq \gamma_G (C(K_{i})-C(K^*)),~\forall i\in \mathbb{Z}_+,
\end{equation}
with $\gamma_G:= 1-\frac{2\eta \lambda_1(\Sigma_w)}{\lVert \Sigma_K^*\rVert}$. For $\eta=\frac{1}{2}$ and given any accuracy gap $\epsilon >0$, if the number of iterations $n\geq \frac{\lVert \Sigma_K^*\rVert}{\lambda_1(\Sigma_w)} \log \frac{C(K_{0})-C(K^*)}{\epsilon}$,
then $C(K_n)-C(K^*) \leq \epsilon.$
\end{thm}
\section{From data to policy gradient via the Bellman equation}
We consider in this work the model-free implementations of NPG and GNM, which directly estimate the quantities appearing in the update rules \eqref{updateNPG} and \eqref{updateGMN}—namely, $B^\top P_{K_i} B$ and $B^\top P_{K_i} A$—without explicitly identifying $(A,B)$. To estimate these matrices, we first collect off-policy data $[x^{(k)},u^{(k)},x^{(k)}_+]^N_{k=1}$ using Algorithm \ref{Algo1}. The required sample size $N$, which determines the estimation error, will be specified later.
\begin{algorithm}[htp]
  \caption{Data Collection}\label{Algo1}
  \begin{algorithmic}
  \For{$k=1,...,N$}
       \State Generate $x^{(k)}\sim \mathcal{N}(0,\Sigma_x)$ and $u^{(k)}\sim\mathcal{N}(0,\Sigma_u)$
        \State Observe the state $x_+^{(k)}$ from the system in \eqref{setting}
      \EndFor
  \end{algorithmic}
\end{algorithm}

After collecting the dataset, for any stabilizing $K$, $B^\top P_{K} B$ and $B^\top P_{K} A$ can be estimated from data. Given an arbitrary random triple $[x,u,x_+]$ from Algorithm \ref{Algo1}, consider the LQR cost in \eqref{setting}, let $u' = Kx$ denote the linear feedback policy, with $x_+'$ the corresponding next state under this policy. The value-based Bellman equation then reads:
\begin{equation}\label{Bell}
\begin{split}
        x^\top P_{K}x+\mathrm{Tr}(P_{K}\Sigma_w)=&\,\,x^\top(Q+K^\top R K)x\\
        &+\mathbb{E}[x_{+}^{'~\top} P_Kx'_{+}|x,K].
\end{split}
\end{equation}
When the system is excited using a generic input $u$ from Algorithm \ref{Algo1} that does not follow the linear policy, we define the deviation $\eta:=u-Kx$. Under this definition we have $x_+'=x_+-B\eta$, \eqref{Bell} can be rewritten as
\begin{equation}
\begin{split}
        x^\top P_{K}x+\mathrm{Tr}(P_{K}\Sigma_w)=x^\top(Q+K^\top R K)x\\
        +\mathbb{E}[(x_+-B\eta)^\top P_K(x_+-B\eta)|x,K,u].
\end{split}
\end{equation}
Replacing $x_+$ with dynamics in \eqref{setting} yields:
\begin{equation}\label{12121212}
\begin{split}
       x ^\top P_{K}x &+\mathrm{Tr}(P_{K}\Sigma_w)=x ^\top(Q+K^\top R K)x \\
       &+\mathbb{E}[x_{+}^\top P_Kx_{+}|x ,K,u ]+\eta ^\top B^\top P_KB \eta \\
       &-2\mathbb{E}[(Ax +BKx +B\eta +w )^\top P_K B \eta ].
\end{split}
\end{equation}
Rearranging terms, the equation can be compactly expressed as
\begin{equation}\label{PIbellmanequation2}
  \begin{split}
\Gamma ^\top \underbrace{\left [\begin{array}{c}
       vec(B^\top P_KA) \\
       vecs(B^\top P_KB)\\
       vecs(P_K)
     \end{array}\right ]}_{=:\xi_K}&=\underbrace{x ^\top (Q+K^\top RK)x }_{=:c },
  \end{split}
\end{equation}
where $\Gamma :=\left [\begin{array}{c}
       2x  \otimes (u -Kx ) \\
       vecv(u )-vecv(Kx ) \\
       vecv(x )+W-\mathbb{E}[vecv(x _+)|x ,K,u ]
     \end{array}\right ]$ 
and $W:=\sum_{k=1}^{n_x}vecv(\sqrt{\lambda_k(\Sigma_w)}v_k)$ with $\lambda_k$ and $v_k$ denoting the $k$-th eigenvalue and eigenvector of the covariance matrix $\Sigma_w$, respectively. The matrix $W$ is introduced so that the trace term in \eqref{12121212} can be expressed as $\mathrm{Tr}(P_K \Sigma_w) = W vecs(P_K)$. Unlike the noise-free case studied in \cite{https://doi.org/10.1002/rnc.7475}, this additional term $W$ appears. 

Since the quantity $\mathbb{E}[\mathrm{vecv}(x_+) \mid x , K, u ]$ is not analytically computable when $A,B$ are unknown. In the model-free setting, we approximate this expectation using the observed sample $x_+$. Accordingly, we define the data-dependent quantity 
\begin{equation}\label{data1}
    \hat{\Gamma} :=\left [\begin{array}{c}
       2x  \otimes (u -Kx ) \\
       vecv(u )-vecv(Kx ) \\
       vecv(x )+W-vecv(x _+)
     \end{array}\right ].
\end{equation}
Given a stabilizing feedback gain $K\in \mathcal{S}$ and dataset $[x^{(k)},u^{(k)},x_+^{(k)}]_{k=1}^N$, we construct the corresponding regression pairs $[\hat{\Gamma}^{(k)}, c^{(k)}]_{k=1}^N$ as defined in~\eqref{data1} and \eqref{PIbellmanequation2}, respectively. The superscript $(k)$ indicates that these quantities are computed from the sample $\{x_t^{(k)}, u_t^{(k)}, x _+^{(k)}\}$. From \eqref{data1}, we have that $\Gamma^{(k)} =\mathbb{E}_{w}[\hat{\Gamma} ^{(k)}],~\forall k\in [1,N]$. Building on the collected dataset, the problem can be formulated as a least-squares (LS) problem as follows:
\begin{equation}\label{els}
    \hat{\xi}_K=\arg \min_{\tilde{\xi}_K}\frac{1}{N}\sum_{k=1}^{N}\lVert \hat{\Gamma}^{(k)} \tilde{\xi}_K-c^{(k)}  \rVert^2.
\end{equation}
However, if the parameter vector $\xi_K$ is estimated using the standard least-squares approach stated in \eqref{els}, then even when the dataset is sufficiently rich such that $\big(\sum_{k=1}^N \hat{\Gamma}^{(k)} \hat{\Gamma}^{(k)\top} \big)\succ 0$, the estimator
\begin{equation}\label{12}
    \hat{\xi}_K=\bigg(\sum_{k=1}^N \hat{\Gamma} ^{(k)}\hat{\Gamma} ^{(k)\top}\bigg)^{-1}\bigg(\sum_{k=1}^N\hat{\Gamma}^{(k)} c ^{(k)}\bigg)
\end{equation}
remains inconsistent. This bias arises from the discrepancy between the random realization $\mathrm{vecv}(x^{(k)} _+)$ and its conditional expectation $\mathbb{E}[\mathrm{vecv}(x^{(k)} _+) \mid x^{(k)}_k, K, u^{(k)}_k]$. Note that $vecv(x^{(k)} _+)$ enters $\hat{\Gamma}^{(k)} $ in \eqref{12}, and thus this is an errors-in-variables problem. To obtain an consistent estimate, we proposed using a primal-dual estimation method, as discussed in the following section. 

\section{Primal-dual Estimation}\label{sec3}

Instead of solving a standard least-squares regression in \eqref{els}, we consider the following \emph{stochastic saddle-point problem} \cite[Section 3.6]{book}:
\begin{equation}\label{111111111}
    \min_{\tilde{\xi}_K\in {X}}\max_{y\in Y}\mathbb{E}_{[x,u,x_+]}[y(\hat{\Gamma} \tilde{\xi}_K-c )],
\end{equation}
where $\hat{\Gamma}$ and $c$ are constructed using the random data triple $[x,u,x_+]$ generated by Algorithm \ref{Algo1}; $Y := \{ y \in \mathbb{R}^1 \mid | y| \le 1 \}$ and $X$ is a compact convex set containing the true parameter $\xi_K$. The construction of the set $X$ will be discussed in Subsection \ref{4.3}. However, we can not directly solve the problem \eqref{111111111} due to expectation. We approximate it using the available $N$ samples, resulting in the \emph{empirical min–max problem}:
\begin{equation}\label{eq:minmax}
    \min_{\tilde{\xi}_K\in {X}}\max_{y\in Y}\bigg\{\frac{1}{N}\sum_{k=1}^Ny(\hat{\Gamma}^{(k)} \tilde{\xi}_K-c^{(k)} )\bigg\}.
\end{equation}
Before presenting the algorithm for solving \eqref{eq:minmax}, we introduce an assumption and several lemmas that will play a key role in the estimation error analysis.
\begin{Ass}[Informativity]\label{ass1}
    Matrix $\Gamma ^{(k)}$ satisfies:
    \begin{equation}
        \Gamma^{(k)~\top} \Gamma^{(k)} \succeq \alpha, \forall k\in [1,N].
    \end{equation}
    for same constant $\alpha>0$.
\end{Ass}
From the expression of $\Gamma ^{(k)}$, the randomness in $[x^{(k)}, u^{(k)}]$ from Algorithm \ref{Algo1} helps ensure that this assumption holds. The reason for introducing this assumption will be explained later. In the stochastic setting, due to the unbounded nature of $[x ^{(k)}, u ^{(k)},x _+^{(k)}]$, the following lemmas establish high-probability bounds on the collected data.

\begin{lem}\label{lem8}
  Let $\delta\in (0,\frac{1}{e}]$ and define the event
  \begin{equation}
  \begin{split}
       \beta^{(k)}(\delta):=\bigg\{ &\lVert [x ^{(k)};u ^{(k)};x _+^{(k)}]\rVert^2 \\
       &\leq4\bigg(\frac{c_2^2}{\sqrt{c_1}}\lVert \tilde{\Sigma} \rVert+\frac{c_2^2}{c_1}\mathrm{Tr}(\tilde{\Sigma})\bigg)\ln \frac{1}{\delta} \bigg\}.
  \end{split}
  \end{equation}
  where $\tilde{\Sigma}:=\left[\begin{array}{ccc}
                                \Sigma_x & 0 & \Sigma_x A^\top \\
                                0 & \Sigma_u & \Sigma_u B^\top \\
                                A\Sigma_x & B\Sigma_u & \Sigma_{x_+} 
                              \end{array}\right]$ with $\Sigma_{x_+}:=A \Sigma_{x}A^\top+B\Sigma_u B^\top+\Sigma_w$ and $c_1,c_2$ are two constants introduced in Lemma \ref{lem7} in Appendix \ref{proof1}. 
  Then, the following holds:
  \begin{equation}\label{11111}
     \mathbb{P}\big[ \beta^{(k)}(\delta)\big]\leq 1-\delta,\forall k\in [1,N].
  \end{equation}  
\end{lem}
The proof of Lemma~\ref{lem8} follows by computing the covariance matrix of the stacked vector $[x ^{(k)}; u ^{(k)}; x _+^{(k)}]$ and applying Lemma~\ref{lem7} from Appendix \ref{proof1}. Consequently, for all $k\in[1,N]$, when $\beta^{(k)}(\delta)$ happens, the associated regression data $\hat{\Gamma} ^{(k)}$ and ${c} ^{(k)}$ are also guaranteed to be bounded.

\begin{lem}\label{Lem9}
Suppose the events $\beta^{(k)}(\delta)$ occur for some $\delta\in (0,\frac{1}{e}]$, for all $k \in [1,N]$. Then, the regression data satisfy
    \begin{equation*}
        \lVert \hat{\Gamma} ^{(k)} \rVert\geq M_{\Gamma}\bigg(\ln\frac{1}{\delta}\bigg),~\lVert c ^{(k)} \rVert\geq M_c\bigg(\ln \frac{1}{\delta}\bigg), \forall k\in[1,N],
    \end{equation*}
    with $M_{\Gamma}$ and $M_c$ defined in \eqref{MH} and \eqref{Mb}, respectively.
\end{lem}
The proof of Lemma \ref{Lem9} is given in Appendix \ref{ProofLemma9}. 
\begin{lem}\label{Lem10}
Let $\bar{\Gamma}:=\mathbb{E}_{[x,u]}[\Gamma ]=\mathbb{E}_{[x,u,x_+]}[\hat{\Gamma} ]$, where $[x,u,x_+]$ is a random data triple from Algorithm \ref{Algo1}. Then, 
\begin{equation*}
\begin{split}    
\lVert \bar{\Gamma} \rVert\leq L_{\Gamma}:=&2\lVert K\rVert \lVert\Sigma_x\rVert_F+\lVert\Sigma_u\rVert
\\&+(\lVert K\rVert^2+1)\lVert \Sigma_x\rVert+\lVert \Sigma_{x_+}\rVert+\lVert W\rVert. 
\end{split}
\end{equation*}
\end{lem}
The proof of Lemma~\ref{Lem10} follows from the definition of $\hat{\Gamma}$ in \eqref{data1}. 
\subsection{Estimation Error Analysis}
We now solve the min–max problem \eqref{eq:minmax} using Algorithm \ref{Algo3} presented below.
\begin{algorithm}[htp]
  \caption{Conditional Stochastic Primal-dual (CSPD)}\label{Algo3}
  \begin{algorithmic}
      \Require Gain matrix $K\in \mathcal{S};X;\xi^{(-1)}_K=\xi_K^{(0)}\in X;N;y^{(0)}\in Y;\{\eta_k,\lambda_k,\zeta_k\}_{k=1}^N$
      \For{$k=1,...,N$}
   \State $G^{(k)}=\xi^{(k-1)}_K+\zeta_k(\xi^{(k-1)}_K-\xi^{(k-2)}_K)$ 
   \State Using the data $\{x^{(k)} ,u^{(k)} ,x^{(k)} _+\}$ and $K$ to construct $\{\hat{\Gamma} ^{(k)},{c} ^{(k)}\}$.  
   \State $y^{(k)}=\arg \min_{\tilde{y}\in Y} \{(-\hat{\Gamma} ^{(k)~\top}G^{(k)}+{c} ^{(k)}) \tilde{y}+\frac{\lambda_k}{2}\lVert \tilde{y}-y^{(k-1)}\rVert^2\}$
   \State $\xi^{(k)}_K=\arg \min_{\tilde{\xi}\in X}\{y^{(k)}(\hat{\Gamma} ^{(k)~\top}\tilde{\xi})+\frac{\eta_k}{2}\lVert \tilde{\xi}-\xi^{(k-1)}_K \rVert^2\}$
      \EndFor 
      \State $(\hat{\xi}_K,\hat{y})=\frac{2}{N(N+1)}\sum_{k=1}^N k\cdot(\xi_K^{(k)},y^{(k)})$
  \end{algorithmic}
\end{algorithm}
Before analyzing the convergence properties of the algorithm, we introduce two key definitions: the \emph{Q-gap function} and the \emph{primal–dual gap}. Let $z' := (\xi'_K, y')$ and $\tilde{z} := (\tilde{\xi}_K, \tilde{y})$. The Q-gap function is defined as
\begin{equation*}
    Q(\tilde{z},z'):=\frac{1}{N}\sum_{k=1}^N(\Gamma ^{(k)}\tilde{\xi}_K-c ^{(k)}) y'-(\Gamma ^{(k)}\xi'_K-c ^{(k)}) \tilde{y},
\end{equation*}
and the primal–dual gap is defined as:
\begin{equation}\label{eq:gap}
    g(\tilde{z}):=\max_{z'\in X,Y}Q(\tilde{z},z').
\end{equation}
Additionally, we introduce the following function:
\begin{equation}\label{f}
    f(\tilde{\xi}_K):=\max_{y\in Y}\left\{\frac{1}{N}\sum_{k=1}^N y(\Gamma ^{(k)}\tilde{\xi}_K-c ^{(k)})\right\}.
\end{equation}
Since $\Gamma ^{(k)} \xi_K = c ^{(k)},\forall k \in [1,N]$, it follows that $g(\hat{z}) \ge f(\hat{\xi}_K)$ where $\hat{z}:=(\hat{\xi}_K,\hat{y})$ is the output of Algorithm \ref{Algo3}. 
The resulting estimation error $\lVert\hat{\xi}_K - \xi_K\rVert$ can be upper-bounded as:
\begin{lem}\label{Lemme9}
    Under Assumption \ref{ass1}:
    \begin{equation}
        \sqrt{\alpha}\lVert \hat{\xi}_K-\xi_K\rVert \leq g(\hat{z})
    \end{equation}
    where $\alpha$ is the constant introduced in Assumption \ref{ass1}.
\end{lem}
\begin{pf}
This result comes directly from the definitions of $f$ in \eqref{f} and set $Y$, which yield:
\begin{equation*}
    g(\hat{z})\geq f(\hat{\xi}_K)=\frac{1}{N}\sum_{k=1}^N\lVert \Gamma ^{(k)}\hat{\xi}_K-c ^{(k)}\rVert\geq \sqrt{\alpha}  \lVert \hat{\xi}_K-\xi_K\rVert. 
\end{equation*}
\end{pf}
Here we emphasize that the functions $Q,g,f$ are defined using $\Gamma^{(k)} $ not the noisy data $\hat{\Gamma} ^{(k)}$. This is precisely why Assumption~\ref{ass1} is stated in terms of $\Gamma ^{(k)}$. Using Lemma~\ref{Lemme9}, to study the estimation error $\Vert\hat{\xi}_K - \xi_K\Vert $, it suffices to analyze the upper bound of the primal–dual gap $g(\hat{z})$. The following theorem provides an upper bound on the Q-gap function:
\begin{thm}\cite[Theorem 3.8]{book}\label{Thm1}
    Let $\{\gamma_k,\eta_k,\lambda_k,\zeta_k\}$ be a set of nonnegative reals satisfying $\gamma_{k-1}\eta_{k-1}\leq \gamma_{k}\eta_{k},\gamma_{k-1}\lambda_{k-1}\leq \gamma_{k}\lambda_{k}$ and $\gamma_k\zeta_k=\gamma_{k-1}$, and let there exist some $p,q\in (0,1)$ satisfying $L_{\Gamma}^2\leq \frac{q\eta_k}{p\lambda_k}$ for all $k\in [1,N]$ and $K\in \mathcal{S}$. Then
    \begin{equation*}
    \begin{split}
        \sum_{k=1}^N\gamma_kQ(z^{(k)},z')\leq  \gamma_N \eta_ND_X^2+\gamma_N\lambda_ND^2_Y+\sum_{k=1}^N\Lambda_k(z'),
    \end{split}
    \end{equation*}
    where $z^{(k)}:=[\xi_K^{(k)},y^{(k)}]$, $D_X^2:=\max_{x_1,x_2\in X}\lVert x_1-x_2\rVert^2$ and $D_Y^2:=\max_{y_1,y_2\in Y}\lVert y_1-y_2\rVert^2$ and 
    \begin{equation}
    \begin{split}
         \Lambda_k(z'):=&-\frac{(1-p)\gamma_k\lambda_k}{2}\lVert y^{(k)}-y^{(k-1)}\rVert^2\\
         &-\frac{(1-q)\gamma_k\eta_k}{2}\lVert \xi_K^{(k)}-\xi_K^{(k-1)}\rVert^2\\
         &+\gamma_k[(\hat{\Gamma} ^{(k)}-\Gamma ^{(k)})G^{(k)}](y^{(k)}-y') \\
         &+\gamma_k[(\hat{\Gamma} ^{(k)}-\Gamma ^{(k)}) y^{(k)}] (\xi_K^{(k)}-\xi'_K).
    \end{split}   
    \end{equation}
\end{thm}
By analyzing the terms $\Lambda_k$, we can derive the following theorem, which bounds the primal–dual gap:

\begin{thm}\label{Thm2}
    Suppose $\beta(\delta):=\cap_{k\in [1,N]} \beta^{(k)}(\delta)$ occurs for some $\delta \in (0,\frac{1}{e}]$. Under the same assumptions as in Theorem \ref{Thm1}, then with probability at least $1-2(N+1)\delta$, 
    \begin{equation*}
    \begin{split}
        \left(\sum_{k=1}^N \gamma_k\right)g(\hat{z})\leq 2(D_XM_X+D_YM_Y)\sqrt{8 \ln \frac{1}{\delta}\sum_{k=1}^{N}\gamma_k^2}\\
        2\gamma_N(\eta_N D_X^2+\lambda_N D_Y^2)
        +\sum_{k=1}^{N}\bigg[\frac{16M_X^2}{\eta_k(1-q)}+\frac{16M_Y^2}{\lambda_k(1-p)}\bigg],
    \end{split}
    \end{equation*}
    where $\hat{z}:=(\hat{\xi}_K,\hat{y})$ is from Algorithm \ref{Algo3}; $\Omega_Y:=\max_{y\in Y}\{\lVert y\rVert\}$; $\Omega_X:=\lVert \xi_K \rVert+(1+\bar{\zeta})\sqrt{2}D_X$; $\bar{\zeta}:=\max_k\zeta_k$; $M_X:=M_{\Gamma}\Omega_Y\ln\frac{1}{\delta}$; $M_Y:=M_{\Gamma} \Omega_X\ln\frac{1}{
    \delta}$.
\end{thm}
The proof of Theorem~\ref{Thm2} is provided in Appendix~\ref{ProofThm2}. By appropriately selecting the parameters ${\eta_k, \zeta_k, \lambda_k}$, we can use Theorem~\ref{Thm2} to derive the estimation error upper bound:
\begin{Corollary}\label{Coro1}
    Let $\eta_k=\frac{3\sqrt{2}L_{\Gamma} D_Y k+6 M_X k^{\frac{3}{2}}}{2\sqrt{2} D_X k}$, $\zeta_k=\frac{k-1}{k}$ and $\lambda_k=\frac{3\sqrt{2}L_{\Gamma}  D_X k+6M_Y k^{\frac{3}{2}}}{2\sqrt{2} D_Y k}$, $\forall k\in [1,N]$. If $\beta(\delta)$ occurs for some $\delta \in (0,\frac{1}{e}]$, under Assumption~\ref{ass1}, then with probability at least $1-2(N+1)\delta$:
    \begin{equation}\label{g}
    \begin{split}
               \lVert \hat{\xi}_K&-\xi_K\rVert\leq \frac{12L_{\Gamma}  D_XD_Y}{\sqrt{\alpha}(N+1)}\\
               &+\frac{2(48+3\sqrt{2}+\frac{16\sqrt{2}}{\sqrt{3}})(D_XM_X+D_YM_Y)}{\sqrt{\alpha}\sqrt{N}}.
    \end{split}
    \end{equation}
\end{Corollary}
\vspace{-5pt}
In view of Corollary~\ref{Coro1}, we observe that the primal–dual optimization method yields the sample complexity $\mathcal{O}(\epsilon^{-2}+\epsilon^{-1})$ for accuracy $\epsilon$ with high probability, and importantly, without bias. 

\subsection{Multi-Epoch Scheme for Reducing Sample Complexity}\label{4.3}
From \eqref{g}, it is evident that if the feasible set $X$ is updated adaptively, the convergence rate can be further improved. Motivated by \cite[Lemma 4.5]{book}, we propose a multi-epoch algorithm that repeatedly invokes Algorithm~\ref{Algo3}, using the solution from the previous epoch to warm-start the next epoch. In each epoch $s$, the initial feasible set for the primal variable, $X_s$, is updated by shrinking it based on the previous epoch, thereby accelerating convergence. The number of samples used in epoch $s$ is denoted by $N_s$. The procedure is summarized in Algorithm~\ref{Algo4}. The following theorem characterizes its convergence properties and the associated sample complexity.
\begin{algorithm}[htp]
  \caption{Shrinking Multi-epoch CSPD}\label{Algo4}
  \begin{algorithmic}
      \Require $K\in \mathcal{S}$; $X$; $\tilde{\xi}_0\in X$; $D_0,S\in \mathbb{R}_{++}$
      \For{$s=1,...,S$}
   \State $D_s^2:=2^{-(s-1)}D_0^2$ 
   \State $X_s:=\{\xi\in X: \lVert \tilde{\xi}_{s-1}-\xi\rVert\leq D_s^2\}$
   \State $\tilde{\xi}_s\leftarrow \hat{\xi}_K$ from Algorithm \ref{Algo3} with $K$; $X_s,~\xi^{(-1)}_K=\xi_K^{(0)}=\tilde{\xi}_{s-1},N_s$ specified by Theorem \ref{TH3}; $y^{(0)}\in Y$; $\{\eta_k,\lambda_k,\zeta_k\}_{k=1}^{N_s}$ specified by Corollary \ref{Coro1}
      \EndFor 
      \State \textbf{Return} $\hat{\xi}_K=\tilde{\xi}_S$
  \end{algorithmic}
\end{algorithm}
\begin{thm} \label{TH3} Consider the same assumptions and parameters as in Corollary \ref{Coro1}. Suppose $\lVert \tilde{\xi}_0-\xi_K \rVert\leq D_0^2$. Define the number of iterations $N_s$ at the $s$-th epoch as
\begin{equation}
    \begin{split}
           N_s:=&\big\lceil 400 \max \big\{ \frac{L_{\Gamma} D_Y}{\alpha},\\
           &\frac{4000+256\ln\frac{1}{\delta}}{\alpha^2}\big(M_X^2+\frac{D_Y^2M_Y^2}{D_0^2}2^s\big)\big\}\big\rceil. 
    \end{split}
\end{equation}
Given an arbitrary accuracy $\epsilon$, choose the number of epochs as $S=\lceil \log_2\frac{D_0^2}{\epsilon} \rceil$. Then, with probability at least $1 - 2(N + S)\delta$, the estimate from Algorithm \ref{Algo4} after $S$ epochs satisfies $\lVert \hat{\xi}_K -\xi_K\rVert\leq \epsilon$ and the total number of iterates over all epochs is at most:
\begin{equation}
\begin{split}
        &N:=400\bigg\lceil \frac{2L_{\Gamma} D_Y}{\alpha}\ln \frac{D_0}{\epsilon}\\
        &+\frac{4000+256 \ln\frac
    {1}{\delta} }{\alpha^2}\bigg(2M_X^2\ln\left(\frac{D_0}{\epsilon}\right)+\frac{D_Y^2M_Y^2}{\epsilon}\bigg)\bigg\rceil.
\end{split}
\end{equation}
\end{thm}
The proof of Theorem~\ref{TH3} follows the steps outlined in \cite[Proposition 4.10]{doi:10.1137/23M1554771}. From Theorem~\ref{TH3}, we can conclude that, to achieve a desired accuracy $\epsilon$, the required sample complexity scales as $\mathcal{O}(\epsilon^{-1}+\ln(\epsilon^{-1}))$. For the initialization of $X_0$, we can select a larger region containing $\xi_K$. Subsequently, Algorithm \ref{Algo4} can be used to shrink the set $X_s$ as the epoch $s$ increases. 

\begin{Rem}
In this section, we proposed to use a primal-dual method to solve min-max linear regression problems with the goal of addressing the bias issue arising in standard least-squares due to errors-in-variables. This contribution is of independent interest and is not limited to solving the Bellman equation \eqref{PIbellmanequation2}; it applies to any regression problem of the form $JL = V$, using data $\hat{J}_k, \hat{V}_k$ that satisfy $\mathbb{E}[\hat{J}_k] = J$ and $\mathbb{E}[\hat{V}_k] = V$ and are bounded with high probability. The primal–dual method yields consistent estimates, achieving a sample complexity of $\mathcal{O}(\epsilon^{-1})$.  
\end{Rem}

\section{Convergence OF THE NPG AND GNM METHODS}
Leveraging the primal–dual optimization procedure described in Section~\ref{sec3}, we are able to estimate the matrices required for implementing the NPG and GNM algorithms to any desired level of accuracy. In this section, we investigate the robustness of the NPG and GNM algorithms with respect to the estimation errors introduced by Algorithm \ref{Algo3} or Algorithm~\ref{Algo4}. The overall procedure is summarized in Algorithm~\ref{Algo5}.
\begin{algorithm}[htp]
  \caption{Model-free NPG/GNM}\label{Algo5}
  \begin{algorithmic}
      \Require $\hat{K}_0\in \mathcal{S}$
      \State Run Algorithm \ref{Algo1} to collect data
      \For{$i=1,...,+\infty$}
      \State Run Algorithm \ref{Algo3} or Algorithm \ref{Algo4} to obtain $\hat{\xi}_{\hat{K}_i}$
      \State Update $\hat{K}_i$ using NPG \eqref{GDS} or GNM \eqref{GDS1}
      \EndFor 
  \end{algorithmic}
\end{algorithm}

The following result provides convergence guarantees for both methods using Algorithm \ref{Algo5}:
\begin{thm}\label{thm:NPG}
    Suppose the initial $\hat{K}_0\in \mathcal{S}$, and consider the natural policy gradient iterates for all $i\in \mathbb{Z}_+$:
    \begin{equation}\label{GDS}
    \hat{K}_{i+1}=\hat{K}_i-2\eta \big[( R + \widehat{B^\top P_{\hat{K}_i} B} ) \hat{K}_i + \widehat{B^\top P_{\hat{K}_i} A}\big],
\end{equation}
where $\widehat{B^\top P_{\hat{K}_i} B},\widehat{B^\top P_{\hat{K}_i} A}$ are the estimates from Algorithm \ref{Algo3} or Algorithm \ref{Algo4} and $\eta\leq \frac{1}{2\lVert R+B^\top P_{\hat{K}_0}B\rVert}$.
Given any accuracy $\epsilon >0 $, $\sigma\in (0,1)$, the number of iterations $n_{\mathrm{N}}$:
\begin{equation*}
    n_{\mathrm{N}}=\frac{\lVert \Sigma_{K^*}\rVert}{2(1-\sigma)\eta \lambda_1(R)\lambda_1(\Sigma_w)}\log \frac{C(\hat{K}_{0})-C(K^*)}{\epsilon}.
\end{equation*}
Given any probability $\delta \in (0,1)$ satisfying $\delta n_{\mathrm{N}}\in (0,1)$, assume that the estimation error of $\hat{\xi}_{\hat{K}_i}$ from Algorithm \ref{Algo3} or Algorithm \ref{Algo4} satisfies, $\forall i\in \mathbb{Z}_+$:
\begin{equation*}
    \mathbb{P}\big\{\lVert \hat{\xi}_{\hat{K}_i} -\xi_{\hat{K}_i}\rVert\leq \frac{\sigma\epsilon\lambda_1{(R)} \lambda_1(\Sigma_w)}{h_C(\hat{K}_0)(1+b_K(C(\hat{K}_0)))\lVert \Sigma_{K^*}\rVert}\big\}\geq 1-\delta,
\end{equation*}
where $h_C,b_K$ were introduced in Lemma \ref{Lipschitz}. Then for any $C(\hat{K}_i)\geq C(K^*)+\epsilon$, the following inequality holds:
\begin{equation*}
        \mathbb{P}\big\{C(\hat{K}_{i+1})-C(K^*) \leq  \hat{\gamma}_{\mathrm{N}} (C(\hat{K}_{i})-C(K^*))\big\}\geq 1-\delta,
\end{equation*}
where $\hat{\gamma}_{\mathrm{N}}:= 1-(1-\sigma)\frac{2\eta \lambda_1{(R)} \lambda_1(\Sigma_w)}{\lVert \Sigma_{K^*}\rVert}< 1$. \\
As a result, the NPG method enjoys the following performance bound:
\begin{equation*}
    \mathbb{P}\big\{ \min_{i\in [0,n_{\mathrm{N}}]}C(\hat{K}_i)-C(K^*)  \leq \epsilon \big\}\geq 1-\delta n_{\mathrm{N}}.
\end{equation*}
\end{thm}
The proof of Theorem \ref{thm:NPG} is provided in Appendix \ref{A4}. Similarly, we can establish the convergence guarantee for the Gauss–Newton method:
\begin{thm}\label{thm:GN}
       Suppose the initial $\hat{K}_0\in \mathcal{S}$, and consider the Gauss-Newton iterates for all $i\in \mathbb{Z}_+$:
    \begin{equation}\label{GDS1}
    \hat{K}_{i+1}=\hat{K}_i-2\eta \left[\hat{K}_i + ( R + \widehat{B^\top P_{\hat{K}_i} B} )^{-1} \widehat{B^\top P_{\hat{K}_i} A}\right],
\end{equation}
where $\widehat{B^\top P_{\hat{K}_i} B},\widehat{B^\top P_{\hat{K}_i} A}$ are the estimates from Algorithm \ref{Algo3} or Algorithm \ref{Algo4} and $\eta\leq \frac{1}{2}$. Given any accuracy $\epsilon >0 $, $\sigma\in (0,1)$, the number of iterations $n_{\mathrm{G}}$:
\begin{equation*}
    n_{\mathrm{G}}=\frac{\lVert \Sigma_{K^*}\rVert}{2(1-\sigma)\eta\lambda_1(\Sigma_w)}\log \frac{C(\hat{K}_{0})-C(K^*)}{\epsilon}.
\end{equation*}
Given any probability $\delta \in (0,1)$ satisfying $\delta n_{\mathrm{G}}\in (0,1)$, assume that the estimation error of $\hat{\xi}_{\hat{K}_i}$ from Algorithm \ref{Algo3} and Algorithm \ref{Algo4} satisfies, $\forall i\in \mathbb{Z}_+$:
\begin{equation*}
    \mathbb{P}\big\{\lVert \hat{\xi}_{\hat{K}_i} -\xi_{\hat{K}_i}\rVert\leq \Delta_{GN}\big\}\geq 1-\delta,
\end{equation*}
where $\Delta_{GD}$ is defined in \eqref{11212}. Then for any $C(\hat{K}_i)\geq C(K^*)+\epsilon$, the following inequality holds:
\begin{equation*}
        \mathbb{P}\big\{C(\hat{K}_{i+1})-C(K^*) \leq  \hat{\gamma}_G (C(\hat{K}_{i})-C(K^*))\big\}\geq 1-\delta,
\end{equation*}
where $\hat{\gamma}_G:= 1-(1-\sigma)\frac{2\eta  \lambda_1(\Sigma_w)}{\lVert \Sigma_{K^*}\rVert}<1.$  \\
As a result, the Gauss-Newton method enjoys the following performance bound:
\begin{equation*}
    \mathbb{P}\big\{ \min_{i\in [0,n_{\mathrm{G}}]}C(\hat{K}_i)-C(K^*)  \leq \epsilon \big\}\geq 1-\delta n_{\mathrm{G}}.
\end{equation*}
\end{thm}
The proof of Theorem \ref{thm:GN} is provided in Appendix \ref{A5}.
For both methods, to achieve a desired accuracy on $C(\hat{K}_i)-C(K^*)$, we can use Algorithms \ref{Algo3} or \ref{Algo4} to ensure that the required precision is met. In particular, to reach a target suboptimality level $\epsilon$, Theorems~\ref{thm:NPG} and \ref{thm:GN} imply that the corresponding estimation error must satisfy $\lVert \hat{\xi}_{\hat{K}_i}-\xi_{\hat{K}_i}\rVert\in\mathcal{O}(\epsilon)$. Since $\lVert \xi_{\hat{K}_i} \rVert \leq (\lVert B\rVert^2+\lVert A\rVert\lVert B\rVert+1)\frac{C(\hat{K}_0)}{\lambda_1(\Sigma_w)}$ and $\lVert \hat{K}_i\rVert\leq b_K(C(\hat{K}_0)),\forall i\in \mathbb{Z}_+$, a unified choice of Algorithm \ref{Algo3} parameters can be made to ensure the estimation error requirement is satisfied at every iteration, according to Corollary \ref{Coro1}. Then, applying Theorem~\ref{TH3} and Corollary \ref{Coro1}, the corresponding sample complexity using Algorithm \ref{Algo4} to achieve this accuracy is therefore $\mathcal{O}(\epsilon^{-1})$.  
The improvement compared with \cite{JMLR:v24:22-0644,doi:10.1137/23M1554771} was achieved thanks to the fact that we reused the collected data in Algorithm \ref{Algo5} instead of having to collect them at each iteration.
\section{Numerical Results}
In this section, we present simulation results\footnote{The Matlab codes used to generate these results are accessible from the repository: {https://github.com/col-tasas/2025-MFPGM-RobustLR}} to illustrate the effectiveness and advantages of Algorithm \ref{Algo5} discussed in the previous sections. We consider the following system, which was already used in prior studies [\cite{9691800,doi:10.1137/23M1554771}]: 
\begin{equation}\label{LTIsimulation}
  x_{t+1}=\underbrace{\left[\begin{array}{ccc}
            1.01 & 0.01 & 0 \\
            0.01 & 1.01 & 0.01 \\
            0 & 0.01 & 1.01 
          \end{array}\right]}_A x_t+\underbrace{\left[\begin{array}{ccc}
            1 & 0 & 0 \\
            0 & 1 & 0 \\
            0 & 0 & 1 
          \end{array}\right]}_B u_t+w_t.
\end{equation}
The weight matrices $Q$ and $R$ are chosen as $0.001I_3$ and $I_3$. The initial gain $\hat{K}_0$ is selected as the optimal LQR solution for $(A,B,100Q,R)$. For data collection, we set $\Sigma_x=\Sigma_u=I_3$ and $\Sigma_w=0.1I_3$. A total of $100$ data are collected. The primal-dual optimization (Algorithm \ref{Algo3}) is configured with $X := \{ \xi \in \mathbb{R}^{21} \mid \lVert \xi\rVert \le 1 \}$, $\zeta_k=\frac{k-1}{k}$, $\lambda_k=\eta_k=0.001\sqrt{k}$. For Algorithm~\ref{Algo4}, we set $S=4$ with epoch sample sizes $N_s=[8,16,24,52]$ and $D_0=1$. For all gradient estimation schemes used to implement the model-free NPG and GNM methods, the step size in \eqref{GDS} and \eqref{GDS1} is fixed at $0.05$. The simulation results are obtained from a Monte Carlo simulation over $30$ data samples. 
\begin{figure}[H]
    \centering
    \includegraphics[width=1\linewidth]{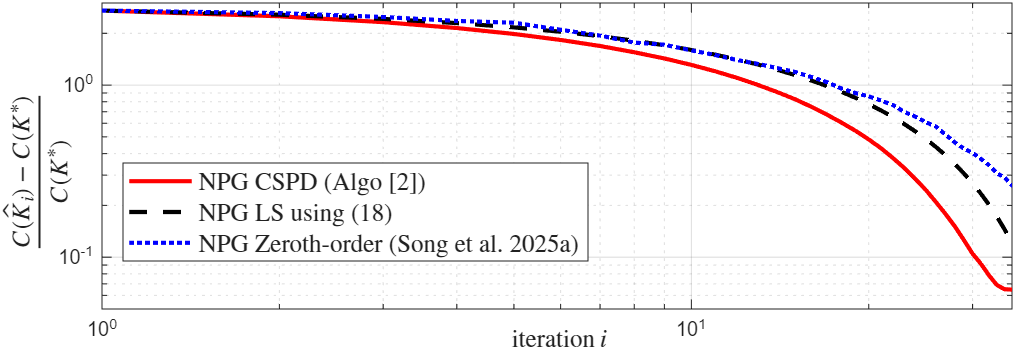}
    \caption{Convergence comparison of model-free NPG using different gradient estimation schemes}
    \label{fig:placeholder}
\end{figure}
\vspace{-10pt}
In Figure \ref{fig:placeholder}, we illustrate the convergence behavior of three model-free NPG methods.
The red solid curve corresponds to the gradient estimated using the CSPD method from Algorithm \ref{Algo3} proposed in Section~\ref{sec3}. We observe that the algorithm converges to the expected suboptimal value. Using the \emph{same dataset}, we also apply the standard least-squares estimator in \eqref{12}, shown as the black dashed curve. As the plot indicates, the convergence based on the gradient estimated via~\eqref{12} is significantly slower than that obtained using the CSPD method. In addition, we compare against the zeroth-order framework by \cite{Full}, plotted as the blue dotted curve. The zeroth-order algorithm \cite[Algorithm 1]{Full} is set to $r=0.4,~n=1000,~l=80$. Implementing this algorithm requires a total of $80*100*35$ samples—substantially more than needed for the CSPD method. The resulting convergence rate of the zeroth-order method is also noticeably slower than that of CSPD.
\begin{figure}[H]
    \centering
    \includegraphics[width=1\linewidth]{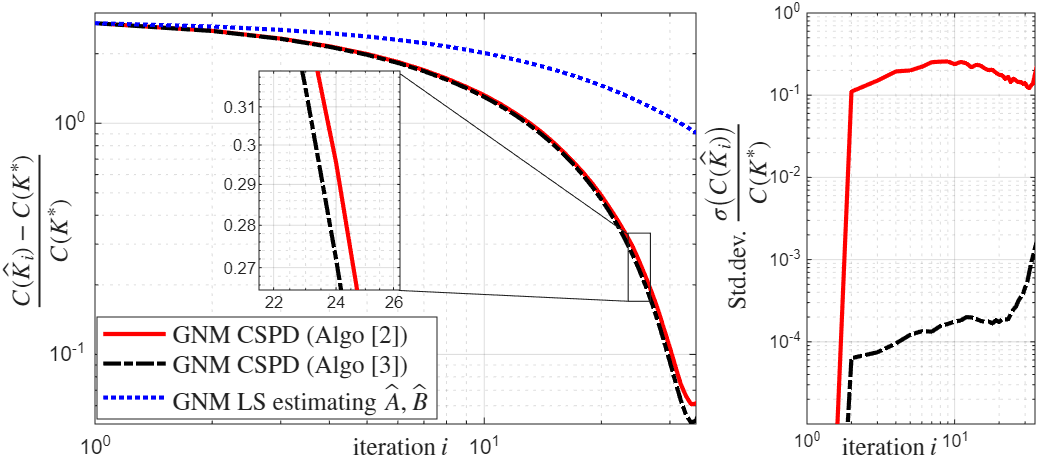}
    \caption{Convergence comparison of model-free GNM using different gradient estimation schemes}
    \label{fig:placeholder2}
\end{figure}
\vspace{-10pt}
In Figure \ref{fig:placeholder2}, we compare the convergence behavior of three model-free GNM methods. The red solid curve corresponds to the gradient estimated using the CSPD method (Algorithm~\ref{Algo3}), while the black dash-dotted curve shows the gradient obtained using the multi-epoch CSPD method (Algorithm~\ref{Algo4}). Both methods use the same dataset, yet the multi-epoch variant achieves faster convergence and a smaller suboptimality gap from the left subplot. The right subplot reports the normalized standard deviation over the Monte Carlo trials, showing that multi-epoch Algorithm \ref{Algo4} also yields smaller variance compared with Algorithm \ref{Algo3}. The blue dotted curve represents the gradient estimation obtained by first identifying the system matrices $\hat{A},\hat{B}$ from the same data and then applying a model-based update, which results in the slowest convergence among the three methods.

\section{Conclusion}
In this work, we employ a primal–dual framework to estimate the required matrices to formulate model-free natural policy gradient and Gauss–Newton methods for the LQR problem using stochastic data. By adopting a multi-epoch scheme and reusing previously collected data, we achieve a sample complexity of $\mathcal{O}(\epsilon^{-1})$ for the convergence of the algorithms, which improves upon the existing results in the literature. Future directions include investigating whether similar convergence guarantees can be established in an online setting, and developing a closed-loop analysis of the overall system that jointly characterizes the dynamics and the algorithmic updates.



\bibliography{ifacconf}             
                                                   







\appendix
\section{Proof}    
\subsection{Lemma \ref{lem7}}\label{proof1}
\begin{lem}\cite[Lemma 4.1]{doi:10.1137/23M1554771}\label{lem7}
  Let $l\sim \mathcal{N}(\mu,\Sigma)$. For any $\delta \in (0,1)$, we have $\mathbb{P}\big\{\frac{1}{2}\lVert l\rVert^2>\mathrm{Tr}(\Sigma)+\frac{\sqrt{c_2^4\lVert \Sigma\rVert_F \log(\frac{2}{\delta})}}{\sqrt{c_1}}+\frac{c_2^2\lVert \Sigma\rVert}{c_1}\log(\frac{2}{\delta})+\lVert \mu\rVert^2\big\}\leq \delta$, where $c_1$ and $c_2$ are some absolute positive constants.
\end{lem}
\subsection{Proof of Lemma \ref{Lem9}}\label{ProofLemma9}
\begin{pf}
    From the definition of $\hat{\Gamma} ^{(k)}$, we have $\lVert \hat{\Gamma} ^{(k)}\rVert\leq [2 \lVert x ^{(k)}\rVert(\lVert u ^{(k)}\rVert+\lVert K\rVert\lVert x ^{(k)}\rVert)+\lVert K\rVert^2\lVert x ^{(k)}\rVert^2+\lVert u ^{(k)}\rVert^2+\lVert x ^{(k)}\rVert^2+ \lVert x_{t+1}^{(k)}\rVert^2+\lVert W \rVert]$, for all $k\in[1,N]$. Using the bound from Lemma \ref{lem8}, we define
    \begin{equation}\label{MH}
    \begin{split}        
    M_{\Gamma}:=4[5+2\lVert K \rVert+\lVert K \rVert^2]\bigg(\frac{c_2^2}{\sqrt{c_1}}\lVert \tilde{\Sigma} \rVert
    +\frac{c_2^2}{c_1}\mathrm{Tr}(\tilde{\Sigma})\bigg)+\lVert W \rVert     
    \end{split}
    \end{equation}
    For $c ^{(k)}$, we have $\lVert c ^{(k)} \rVert\leq (\lVert Q\rVert+\lVert K\rVert^2\lVert R\rVert)\lVert x ^{(k)}\rVert^2$, then
    \begin{equation}\label{Mb}
        M_c:=4(\lVert Q\rVert+\lVert K\rVert^2\lVert R\rVert)\bigg(\frac{c_2^2}{\sqrt{c_1}}\lVert \tilde{\Sigma} \rVert+\frac{c_2^2}{c_1}\mathrm{Tr}(\tilde{\Sigma})\bigg).
    \end{equation}
\end{pf}

\subsection{Proof of Theorem \ref{Thm2}}\label{ProofThm2}
At first, we introduce the following lemma:
\begin{lem}\label{Lemma3}
    Let $\{\gamma_k,c_k\}$ be nonnegative scalars satisfying $\gamma_{k-1}c_{k-1}\leq \gamma_{k}c_{k}$. Let $G_k$ be a stochastic vector such that 
    \begin{equation}\label{error1}
        \mathbb{P}[\lVert G_k-\mathbb{E}[G_k]\rVert  \leq \epsilon]\geq 1-\delta 
    \end{equation}
    Then, for any vectors $u_k,v_k,u',v'\in U$, with probability at least $1-(N+1)\delta$,
    \vspace{-5pt}
    \begin{equation*}
    \begin{split}
       \sum_{k=1}^N\gamma_k (G_k-\mathbb{E}[G_k])^\top (u_k-u')-\frac{c_k\gamma_k}{2}\lVert u_k-u_{k-1}\rVert^2\\
       \leq \gamma_Nc_ND_U^2+\sum_{k=1}^N\frac{\gamma_k}{c_k}\epsilon^2 +\epsilon D_U\sqrt{8\ln \epsilon^{-1}\sum_{k=1}^N \gamma_k^2}. 
    \end{split}
    \end{equation*}
    where $D_U := \max_{u_1,u_2\in U} \lVert u_1 - u_2 \rVert$.
\end{lem}
\begin{pf}
    Define $u_0^v:=u_0,\Delta_k:=G_k-\mathbb{E}[G_k]$ and let $u_k^v:=\arg\min_{u\in U} \Delta_k^\top u_k+\frac{c_k}{2}\lVert u-u_{k-1}^v\rVert^2$. We decompose the sum as: $\sum_{k=1}^N\gamma_k\Delta_k^\top(u_k-u')-\frac{c_k\gamma_k}{2}\lVert u_k-u_{k-1}\rVert^2=\sum_{k=1}^NA_k+B_k+C_k$
    with $A_k:=\gamma_k\Delta_k^\top(u_k-u_{k-1})-\frac{c_k\gamma_k}{2}\lVert u_k-u_{k-1}\rVert^2$ and $B_k:=\gamma_k\Delta_k^\top(u_k-u^v_{k-1})$ and $C_k:=\gamma_k\Delta_k^\top(u^v_{k-1}-u')$.
    We bound each term separately: Using Young’s inequality and \eqref{error1}, with probability at least $1 - N\delta$,
    \begin{equation*}
        \sum_{k=1}^NA_k\leq \sum_{k=1}^N\frac{\gamma_k}{2c_k}\lVert \Delta_k \rVert^2\leq \sum_{k=1}^N\frac{\gamma_k}{2c_k}\epsilon^2. 
    \end{equation*}
    Applying the Azuma–Hoeffding inequality and the bound on $\Delta_k$, with probability at least $1 - \delta$,
    \begin{equation*}
    \begin{split}
        \sum_{k=1}^NB_k\leq \sum_{k=1}^N(B_k-\mathbb{E}(B_k))+\mathbb{E}(B_k)\leq \epsilon D_U \bigg[8\ln\frac{1}{\delta}\sum_{k=1}^N\gamma_k^2\bigg]^{\frac{1}{2}}
    \end{split}
    \end{equation*}
    Using $\gamma_{k-1}c_{k-1}\leq \gamma_{k}c_{k}$, we have
    \begin{equation*}
        \sum_{k=1}^NC_k\leq \gamma_Nc_ND_U^2+\sum_{k=1}^N\frac{\gamma_k}{c_k}\lVert \Delta_k\rVert^2\leq \gamma_Nc_ND_U^2+\sum_{k=1}^N\frac{\gamma_k}{2c_k}\epsilon^2
    \end{equation*}
    Combining the bounds for $A_k$, $B_k$, $C_k$ completes the proof.
\end{pf}
Now we start the proof of Theorem \ref{Thm2}
\begin{pf}
  We start by decomposing $\sum_{k=1}^N \Lambda_k(z')$ in Theorem \ref{Thm1} as: $\sum_{k=1}^N \Lambda_k(z')\leq \sum_{k=1}^N-\frac{(1-p)\gamma_k\lambda_k}{2}\lVert y^{(k)}-y^{(k-1)}\rVert^2-\frac{(1-q)\gamma_{k}\eta_k}{2}\lVert \xi_K^{(k)}-\xi_K^{(k-1)}\rVert^2+\gamma_k[(\hat{\Gamma}_t^{(k)\top}g^{(k)})-(\Gamma ^{(k)\top}g^{(k)})]^\top(y^{(k)}-y') +\gamma_k[(\hat{\Gamma}^{(k)\top}_ky^{(k)}-\Gamma^{(k)\top}  y^{(k)})]^\top (\xi_K^{(k)}-\xi'_K)$. To bound these terms, we apply Lemma \ref{Lemma3} separately: the first and third terms correspond to the stochastic error in the dual variable $y^{(k)}$, involving the vector $(\hat{\Gamma} ^{(k)\top} g^{(k)} - \Gamma ^{(k)\top} g^{(k)})$; the second and fourth terms correspond to the stochastic error in the primal variable $\xi_K^{(e)}$, involving the vector $(\hat{\Gamma}^{(k)\top}  y^{(k)} - \Gamma ^{(k)\top} y^{(k)})$. To apply Lemma \ref{Lemma3}, we need bounds on the norms of these stochastic vectors. Using Lemma \ref{Lem9}, we can compute high-probability upper bounds for both $\hat{\Gamma} ^{(k)\top} g^{(k)}$ and $\hat{\Gamma} ^{(k)\top} y^{(k)}$. The rest of the proof then follows the steps in \cite[Theorem 4.8]{doi:10.1137/23M1554771}.
\end{pf}

\subsection{Proof of Theorem \ref{thm:NPG}}\label{A4}
\begin{pf}
We denote the update with exact parameters as $\bar{K}_i:=\hat{K}_i-2\eta [( R + {B^\top P_{\hat{K}_i} B} ) \hat{K}_i + {B^\top P_{\hat{K}_i} A}]$. From Theorem \ref{Thm1}, we know that $C(\bar{K}_{i+1})-C(K^*)\leq \gamma_N (C(\hat{K}_i)-C(K^*))$. If $|C(\hat{K}_{i+1})-C(\bar{K}_{i+1})|\leq \frac{2\sigma\epsilon\eta \lambda_1{(R)} \lambda_1(\Sigma_w)}{\lVert \Sigma_{K^*}\rVert}$, then we have $ C(\hat{K}_{i+1})-C(K^*)\leq \hat{\gamma}_N(C(\hat{K}_i)-C(K^*))$. From the definition of $\xi_K$, we know that $\lVert\hat{\xi}_K- \xi_K\rVert=\lVert\hat{\xi}_K- \xi_K\rVert_F\geq \lVert B^\top {P_K}B\rVert, \lVert B^\top{P_K}A\rVert$. Comparing the updates with exact and estimated parameters, we have $\lVert \bar{K}_i-\hat{K}_i\rVert\leq2\eta (1+\lVert \hat{K}_i\rVert)\lVert \tilde{\xi}_{\hat{K}_i}-\xi_{\hat{K}_i}\rVert$. Using the Lipschitz continuity of $C(K)$, it follows that if $\lVert\tilde{\xi}_{\hat{K}_i}-\xi_{\hat{K}_i}\rVert\leq \frac{\sigma \epsilon\lambda_1{(R)} \lambda_1(\Sigma_w)}{h_C(\hat{K}_0)(1+b_K(C(\hat{K}_0)))\lVert \Sigma_{K^*}\rVert},$ contraction is guaranteed at each iteration. Further, we have $\lVert \hat{K}_i\rVert\leq b_K(C(\hat{K}_0)),\forall i\in\mathbb{Z}_+$. Applying this analysis iteratively and using a union bound, we obtain the overall convergence result. The rest detailed steps can be found in \cite{Full}.
 
\end{pf}
\subsection{Proof of Theorem \ref{thm:GN}}\label{A5}
\begin{pf}
    Similarly to the previous proof, we denote $\bar{K}_i$ as the update using exact matrices: $ \bar{K}_{i+1}=\hat{K}_i-2\eta [\hat{K}_i + ( R + {B^\top P_{\hat{K}_i} B} )^{-1} {B^\top P_{\hat{K}_i} A}],$ From Theorem \ref{Thm2}, we have $ C(\bar{K}_{i+1})-C(K^*)\leq \gamma_G (C(\hat{K}_i)-C(K^*))$.
    If $|C(\bar{K}_i)-C(\hat{K}_i)|\leq \frac{2\sigma\eta\lambda_1(\Sigma_w)}{\lVert \Sigma_K^*\rVert}$, then $C(\hat{K}_{i+1})-C(K^*) \leq  \hat{\gamma}_G (C(\hat{K}_{i})-C(K^*))$. By the Lipschitz continuity of $C$, we define $\Delta_G:=\lVert  ( R + \widehat{B^\top P_{\hat{K}_i} B} )^{-1} \widehat{B^\top P_{\hat{K}_i} A}- ( R + {B^\top P_{\hat{K}_i} B} )^{-1} {B^\top P_{\hat{K}_i} A}\rVert $ and require $\Delta_G\leq \frac{\sigma\epsilon\lambda_1(\Sigma_w)}{h_C(C(\hat{K}_0))\lVert \Sigma_K^*\rVert}$. We can bound $\Delta_G$ as $\Delta_G\leq \lVert (R+\widehat{B^\top P_{\hat{K}_i}B})^{-1}-(R+B^\top P_{\hat{K}_i} B)^{-1}\rVert\lVert {B^\top P_{\hat{K}_i} A}\rVert+\lVert {B^\top P_{\hat{K}_i} A-\widehat{B^\top P_{\hat{K}_i} A}}\rVert\lVert (R+\widehat{B^\top P_{\hat{K}_i}B})^{-1}\rVert$. For the first term, since $\lambda_1(R+B^\top P_{\hat{K}_i}B)\geq \lambda_1(R)$, using the perturbation bound for matrix inverses \cite[Theorem 33]{pmlr-v80-fazel18a}, if $\lVert \hat{\xi}_{\hat{K}_i}-\xi_{\hat{K}_i}\rVert\leq \frac{\lambda_1(R)}{2}$, then $\lVert (R+\widehat{B^\top P_{\hat{K}_i}B})^{-1}-(R+B^\top P_{\hat{K}_i} B)^{-1}\rVert\leq 2\lVert \hat{\xi}_{\hat{K}_i}-\xi_{\hat{K}_i}\rVert/\lambda_1(R)^2\leq \lVert \hat{\xi}_{\hat{K}_i}-\xi_{\hat{K}_i}\rVert $. The second term is upper bounded by $\lVert \hat{\xi}_{\hat{K}_i}-\xi_{\hat{K}_i}\rVert(\lVert R^{-1}\rVert+\frac{\lambda_1(R)}{2})$. Summarizing both terms, we have $\Delta_G\leq \bar{\Delta}_G\lVert \hat{\xi}_{\hat{K}_i}-\xi_{\hat{K}_i}\rVert$ with $\bar{\Delta}_G:=(\lVert R^{-1}\rVert+\frac{\lambda_1(R)}{2}+\frac{\lVert A\rVert\lVert B\rVert C(\hat{K}_0)}{\lambda_1(\Sigma_w)})$. To guarantee contraction at each iteration, it suffices to enforce $\lVert \hat{\xi}_{\hat{K}_i}-\xi_{\hat{K}_i}\rVert\leq \Delta_{GN}$, with
    \begin{equation}\label{11212}
    \begin{split}
        \Delta_{GN}:=\min\bigg(\frac{\lambda_1(R)}{2},\frac{\sigma\epsilon\lambda_1(\Sigma_w)}{h_C(C(\hat{K}_0))\lVert \Sigma_{K^*}\rVert\bar{\Delta}_G}\bigg).
    \end{split}
    \end{equation}
Applying this bound iteratively completes the proof.
\end{pf}

\end{document}